**Copper and silver thin film systems display differences in antiviral and antibacterial properties – implications for the prevention of SARS-CoV-2 infections**

*Toni Luise Meister[1,]\*, Jill Fortmann[2,]\*, Marina Breisch[3,]\*, Christina Sengstock[3,4], Eike Steinmann[1], Manfred Köller[3], Stephanie Pfaender,[1#] Alfred Ludwig[2,#]*

[1] Department for Molecular and Medical Virology, Ruhr University Bochum, Universitätsstraße 150, 44801 Bochum, Germany
stephanie.pfaender@rub.de

[2] Chair for Materials Discovery and Interfaces, Institute for Materials, Ruhr University Bochum, Universitätsstraße 150, 44801 Bochum, Germany
alfred.ludwig@rub.de

[3] BG University Hospital Bergmannsheil Bochum / Surgical Research, Ruhr University Bochum, Buerkle-de-la-Camp-Platz 1, 44789 Bochum, Germany

[4] Current address: Leibniz-Institut für Analytische Wissenschaften - ISAS – e.V., Bunsen-Kirchhoff-Straße 11, 44139 Dortmund, Germany.

\* These authors contributed equally, [#] corresponding authors

**Abstract**

The current Coronavirus Disease 19 (COVID-19) pandemic has exemplified the need for simple and efficient prevention strategies that can be rapidly implemented to mitigate infection risks. Various surfaces have a long history of antimicrobial properties and are well described for the prevention of bacterial infections. However, their effect on many viruses has not been studied in depth. In the context of COVID-19, several surfaces, including copper (Cu) and silver (Ag) coatings have been described as efficient antiviral measures that can easily be implemented to slow viral transmission. In this study, we detected antiviral properties against Severe Acute Respiratory Syndrome Coronavirus-2 (SARS-CoV-2) on surfaces, which were coated with Cu by magnetron sputtering. However, no effect of Ag on viral titers was observed, in clear contrast to its well-known antibacterial properties. Further enhancement of Ag ion release kinetics based on an electrochemical sacrificial anode mechanism did not increase antiviral activity. These results clearly demonstrate that Cu and Ag thin film systems display significant differences in antiviral and antibacterial properties which need to be considered upon implementation.

**1. Introduction**

Cu and Ag are known as antimicrobial agents for centuries, however, in the medical field these metals have experienced a renaissance over the last years due to the increasing emergence of antibiotic-resistant microorganisms. Beside applications of these metals in numerous consumer products they are used in various biomaterials or healthcare settings to prevent bacterial colonization of implants and devices or to support hospital hygiene procedures to reduce hospital-acquired infections. Both, Cu and Ag, exert broad antimicrobial activities (bacteria, fungi and viruses) and show a low incidence to induce microbial resistance as both attack a broad range of targets in microorganisms.[1–3] Especially the pandemic spread of Severe Acute Respiratory Syndrome Coronavirus-2 (SARS-CoV-2) causing the disease COVID-19 has exemplified the requirement for effective public health intervention strategies that contribute to controlling virus transmission. However, the development of antiviral surfaces which are able to inactivate adherent virus particles and thereby hinder virus transmission from contaminated surfaces is challenging due to the different inherent properties of microbes compared to viruses. The antibacterial activity of Ag is strongly related to the release of Ag ions ($Ag^+$) which are formed by oxidative dissolution, while in contrast, zero valent Ag ($Ag^0$) exerts no comparable antibacterial activity. [2,4–6] $Ag^+$ ions interact with a variety of biomolecules within a cell such as cell membrane and cell wall components, thiol ligands, e.g., sulfhydryl groups of metabolic enzymes, or nucleic acids, and others. Furthermore, reactive oxygen species (ROS) are generated due to $Ag^+$ ions which leads to harmful oxidative stress effects.[2,7,8] In general, consequences are biomolecule damage and subsequent cellular dysfunctions which finally inhibit bacterial proliferation up to bactericidal effects. The antibacterial efficiency of Ag can be enhanced by an increase in the $Ag^+$ releasing surface area by using e.g., Ag nanoparticles.[9] In addition, recently, we presented a concept to enhance $Ag^+$ release kinetics based on an electrochemical sacrificial anode mechanism.[10–12] By combination of Ag with a more noble metal (Au, Pt, Pd, or Ir) within an electrolytic environment (such as biological fluids) the less noble Ag corrodes in favor of the more noble part (it is "sacrificed"). We have demonstrated that such sacrificial anode surfaces exert much higher antibacterial effects compared to pure Ag surfaces with much higher total $Ag^+$ due to the electrochemically driven enhanced dissolution of Ag. Based on these results we aimed at analyzing the antiviral properties of surfaces coated with Cu or Ag as well as several Ag-based sacrificial anode surfaces including combinations of Cu and Ag for possible synergistic effects and compare antiviral against antimicrobial performance.

## 2. Results

The results were achieved using different sputtered thin film surfaces (for details see *Experimental Methods*): (I) continuous and dense thin films of Ag and Cu (thickness 50 nm) and (II) nanostructured surfaces with high surface areas. The latter were synthesized by (a) sequentially depositing Ag on Pt, Cu on Ag or (b) by co-sputtering Ag and Pt as well as Cu and Ag. We call the resulting surface structures "nanopatches", i.e., nanoislands which are formed by the two metals. In case of sequential sputtering (Ag on Pt, Cu on Ag) the elements tend to be more separated as compared to co-sputtering which tends to mix the elements on the atomic scale (Ag&Pt, Cu&Ag) and rather forms an alloy (forced solid solution) compared to the co-existence of elemental films (Ag on Pt, Cu on Ag). The latter is expected to have better sacrificial anode properties. We synthesized nanopatches of two different thicknesses, which offer different volumes of material (e.g., thin Ag/Pt vs. thick Ag/Pt) for releasing metal ions (Table 1).

In order to qualitatively compare antimicrobial and antiviral properties of sputtered Ag and Cu surfaces, we first evaluated the antibacterial properties of thin Ag/Pt and Ag/Cu sacrificial anode nanopatches. Bacterial tests were performed with *Staphylococcus aureus* (*S. aureus*) using a drop-based experimental setup allowing analysis of planktonic and adherent bacteria. Planktonic bacteria within the drop were quantified by plating on blood agar plates, while adherent bacteria on the sample surface were visualized by fluorescence microscopy.

Pure Ti thin films as well as thin nanopatches of pure Pt, Ag, and Cu served as controls and exhibited no significant antibacterial activity against *S. aureus* (**Figure 1A**). Similarly, bacterial growth was not affected by co-deposited thin Ag/Pt nanopatches indicating the absence of a sacrificial anode effect (**Figure 1B**). In contrast, sequentially-deposited thin Ag/Pt nanopatches as well as thin Ag/Cu nanopatches both co- and sequentially-deposited, effectively prevented bacterial growth after 24 h of incubation (**Figure 1B**).

As it is generally accepted that the antibacterial activity of Ag und Cu is strongly related to the release of ions and their interaction with cellular components and processes, solutions of silver acetate (AgAc) and copper sulfate (CuSO$_4$) were used as ionic controls for the antibacterial activity of Ag and Cu towards the gram positive bacteria *S. aureus*.[13,14] Significant antibacterial effects were detected for AgAc at concentrations ≥ 1.0 µg/mL, whereas CuSO$_4$ provoked significant effects starting at concentrations ≥ 5.0 µg/mL (**Figure 2**).

These results indicate that the absence of antibacterial effects of pure Ag and Cu nanopatches are due to an insufficient ion release from these structures, while the combination of Ag and Pt as well as Ag and Cu leads to enhanced antibacterial activity based on electrochemically driven enhanced dissolution of Ag and Cu, respectively (**Figure 1**). Previously, we demonstrated such sacrificial anode effects for nanoparticular and nanostructured Ag/Pt systems.[10,12,15] Regarding the Ag/Cu system, in addition to a possible sacrificial anode effect of Ag on Cu, a combination effect of the two antibacterial metals might be considered. [10,12]

After demonstrating antibacterial effects of sputtered Ag and Cu surfaces as well as ionic Ag and Cu solutions, in accordance with previous results, we aimed in a next step at analyzing potential antiviral effects of these surfaces against SARS-CoV-2. Viral contamination was mimicked upon inoculation of SARS-CoV-2 onto surfaces for either 1 h or 24 h, before viral titers were determined as TCID$_{50}$/mL (**Figure 3**). While Ag nanopatches did not affect viral infectivity, nanopatches with a thick layer of Cu reduced viral titers by 1 log$_{10}$ when incubated together with SARS-CoV-2 for 24 h (**Figure 3A**). An even more pronounced antiviral efficacy was observed for thin Cu films sputtered on Si/SiO$_2$ pieces after 1 h and 24 h of incubation reducing viral titers for 3 log$_{10}$ and 4.5 log$_{10}$, respectively, whereas Ag films did not reduce viral titers (**Figure 3B**). Interestingly, Cu&Ag and Cu on Ag nanopatches reversed the antiviral properties of solely Cu, with almost no antiviral effect after 1 h and reduced antiviral activity after 24 h incubation. While thin films reduced viral titers by 2.2 log$_{10}$, thick films increased virus inactivation resulting in reduction factors of 3.9 log$_{10}$ (**Figure 3C**). This demonstrates that pure Cu films offer the highest antiviral effect (**Figure 3B**). In contrast, co-deposited and sequentially deposited Ag-Pt nanopatches did not reduce viral infectivity within 1 h incubation. A mild antiviral effect (1 $\log_{10}$ reduction of viral titers) was observed after 24 h incubation with thin Ag&Pt and thin Ag on Pt nanopatches (**Figure 3D**). In order to explain the observation that Ag-coated surfaces do not affect viral infectivity, we used silver acetate solutions (AgAc) as a reference to test antiviral properties of Ag at higher ion concentrations as compared to what can be released from a thin film surface. We included concentrations ranging from 1 µg/mL up to 50 µg/mL Ag and inoculated the solution with SARS-CoV-2 containing supernatant for 1 h or 24 h before determining viral infectivity as $TCID_{50}$/mL. Only concentrations equal to or higher than 25 µg/mL displayed antiviral properties and completely abolished infectivity of SARS-CoV-2, however, only upon prolonged incubation of 24 h (**Figure 3E**). In contrast, virus exposure towards $CuSO_4$ solution in a similar experimental setup had no effect on viral titers (**Figure 3F**). In conclusion, we demonstrate a clear antiviral effect of Cu-coated surfaces against SARS-CoV-2 within 1 h exposure, whereas Ag did not influence viral infectivity.

## 3. Discussion

Rapid and effective prevention measures are urgently needed in order to combat microbial and viral diseases. Fomite transmission via contaminated surfaces was described for a variety of microbes, including several viruses and has been discussed in the context of COVID-19. Even though surfaces are believed to play a minor role for the spread of SARS-CoV-2, public health intervention strategies still rely on disinfectant procedures in order to reduce viral transmission. Various surface coatings exhibiting topographically and/or chemically induced antimicrobial activity have been suggested, including nanostructured materials as well as materials containing antimicrobial agents (e.g. antibiotics, antiviral drugs, nanoparticles) such as commercially available antibacterial/antiviral foils, textiles, paints, and many more.[16,17] Although disinfection can be effective in the prevention of infection spread, the biologically active agents of many widely used surface and hand disinfectants might also be hazardous to humans and the environment, especially at prolonged application or misuse. In particular, skin and ocular irritation as well as chemical burns to the respiratory track might occur. Disruption of the normal skin flora, which normally represents a protection barrier to harmful agents, can even enhance the risk of infection.[18]

In line with van Doremalen et al., SARS-CoV-2 was less stable on sputtered Cu surfaces compared to all other thin films of this study, although partially detectable after 24 h of incubation.[19] A thicker film led to a more pronounced antiviral effect. However, combining Cu with Ag by either co-sputtering or sequentially depositing Cu on Ag, reduced antiviral effects monitored by limited dilution assays. Thin films of pure Ag or Ag-based sacrificial anode nanopatches (Ag in combination with Pt) displayed no capacity to inactivate SARS-CoV-2 (Figure 3 B-D). Only silver acetate solutions with concentrations ≥ 25 µg/mL resulted in a 4-log reduction of viral titers when incubated for 24 h. Reducing the silver acetate concentration or incubation period completely abolished any antiviral effects (Figure 3E). This means that only very high concentrations of Ag ions have an effect on SARS-CoV-2.

These results clearly indicate differences between antiviral and antibacterial activities of Ag. For inactivation of SARS-CoV-2 about 10-fold higher concentrations of Ag ions are required compared to efficient antibacterial concentrations. The reason may be related to the difference in the nature of viruses vs. bacteria. As living organisms, bacteria offer much more Ag-sensitive metabolic processes such as energy generation or cell proliferation compared to a virus particle. Furthermore, some bacteria such as *S. aureus* exhibit a significant inherent tolerance of growth at high Cu concentrations.[20] However, a copper sulfate solution up to 50 µg/mL did not inactivate SARS-CoV-2 (Figure 3F). In our study the ionic Ag and Cu solutions had similar concentration up to 50 µg metal/mL to allow direct comparison. Whereas ionic Ag is antibacterial at concentrations of ≥ 1 µg/mL, ionic Cu is required at much higher concentrations to inhibit bacterial growth (e.g. 10 mM $CuSO_4$).[20,21] Huang et al. (2008) observed similar differences in antibacterial efficiency between Ag and Cu ions, however, at lower total ion concentrations due to the water test medium.[22] Thus, at equimolar concentrations Ag ions shows better antibacterial performance compared to Cu ions. Virus inhibition by soluble Cu ions might require higher concentrations above the used maximal values (50 µg/mL) in our experiments.

In contrast to ionic species, antibacterial and antiviral activities of solid Ag and Cu sputtered surfaces have led to different results. Nanopatches of pure Ag neither exhibit antibacterial nor antiviral activities. Obviously, Ag ion release is insufficient under these experimental conditions. To overcome this limited Ag ion release, the more active ion-releasing sacrificial anode surfaces can be used as they exhibit much better antibacterial efficiency even at lower total Ag content [10,12] as was also observed here for Ag/Pt samples (Figure 1). However, even these antibacterial Ag/Pt samples failed to induce any antiviral effect which indicates again the need of higher Ag ion concentrations to reach antiviral activity.

Remarkably, our study shows that solid-state Cu either as a dense film or as nanopatches is able to induce antiviral activity, but not solid-state Ag. The antiviral effects are dependent on the total Cu amount (thickness of the sputtered Cu) and on the time of sample exposure.

It was reported that solid-state cuprous compounds exhibit efficient antiviral activities, whereas those of solid-state Ag are markedly lower.[23] In particular the inactivation of influenza virus HA and NA surface proteins are affected by the exposure to Ag and Cu.[24] Solid-state Ag is less susceptible than solid-state Cu to surface oxidation under the experimental conditions to release ionic species.[25] It is known that several metabolic products of Cu such as cuprous oxide ($Cu_2O$), sulfide ($Cu_2S$), or chloride ($CuCl$) exhibit high antiviral activities and Cu surfaces retain their anti-infective properties even after oxide formation.[23,26]

Although there are numerous reports on antibacterial and antiviral effects of Ag or Cu and their related compounds, a direct comparison of Ag and Cu is rarely found and a high variability of the reported study methods makes such direct comparison difficult.[27]

Taken together, biocidal surfaces could provide constant antiviral and antibacterial efficacy against reoccurring contamination, thus reducing the spread of certain pathogens, given that the surface stays clean and is not used up, whereas surface disinfection has to be reapplied with every contamination.[28] The antimicrobial activity of Cu-based materials and surfaces was demonstrated against different pathogens, including SARS-CoV-2, MRSA (meticillin-resistant *S. aureus*), VRE (vancomycin resistant enterococci), and other nosocomial microbes, while techniques such as cold-spray coating or Cu-impregnation would circumvent the need to completely replace existing surfaces.[29–32] However, incubation periods greater than 1 h are not applicable for many administrations and prevention measures should therefore be critically evaluated with respect to the targeted pathogen.

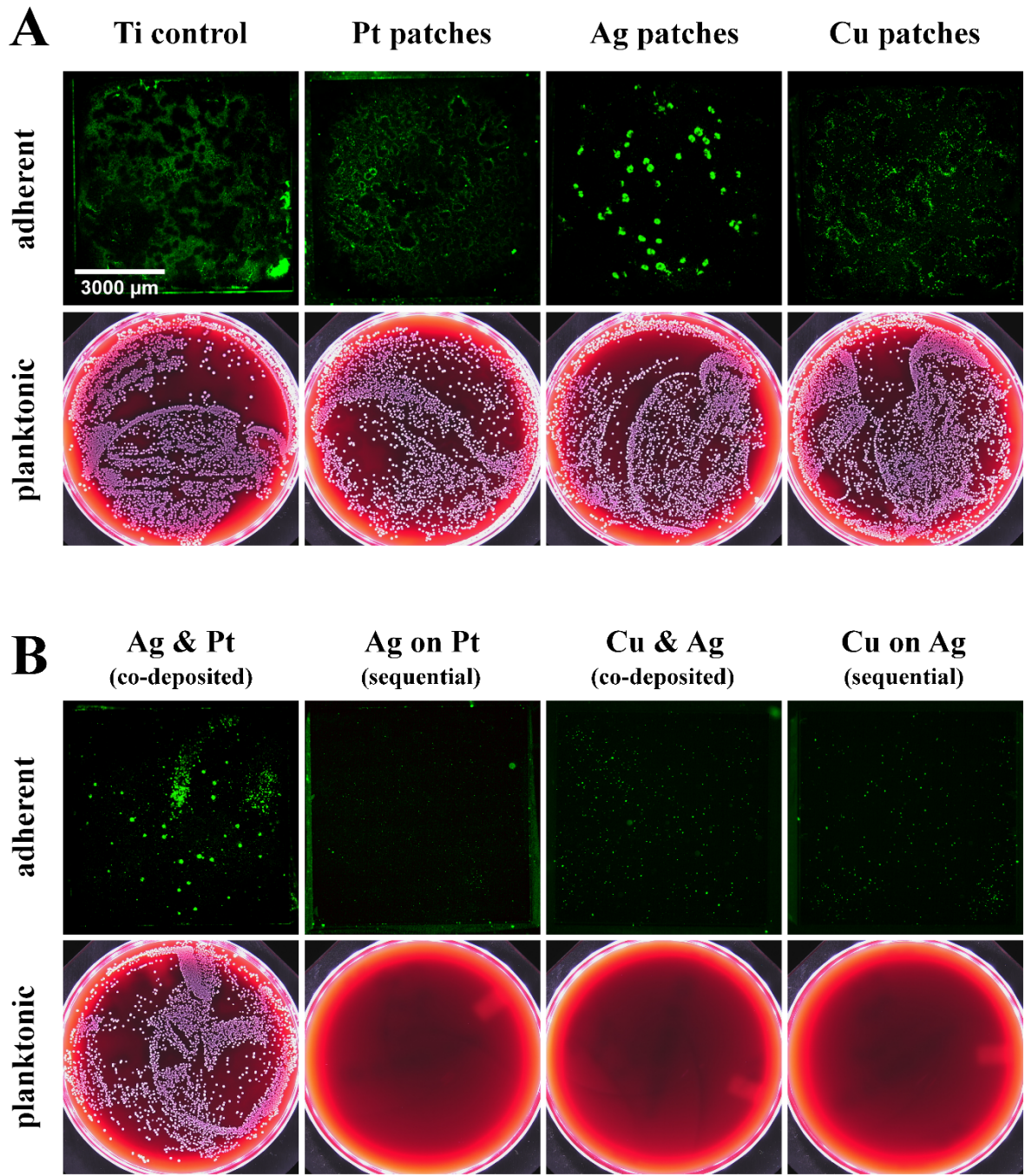

**Figure 1**: Antibacterial activity towards *S. aureus* ($10^4$ CFU/mL) of (A) a continuous Ti thin film (Ti control) as well as Pt, Ag and Cu thin nanopatches sputtered on Ti thin film compared to (B) thin Ag/Pt and thin Ag/Cu nanopatches sputtered simultaneously (i.e., co-deposited) or sequentially (first Pt, second Ag or first Ag, second Cu). Sputter time for all samples 60 s. Upper figures: representative fluorescence images of adherent bacteria on sample surfaces after 24 h of incubation and staining with SYTO-9 (green fluorescence); lower images: representative blood agar plates of plated of planktonic bacteria in the drop fluid after 24 h of incubation on the different samples (white bacterial colonies indicate viable cells).

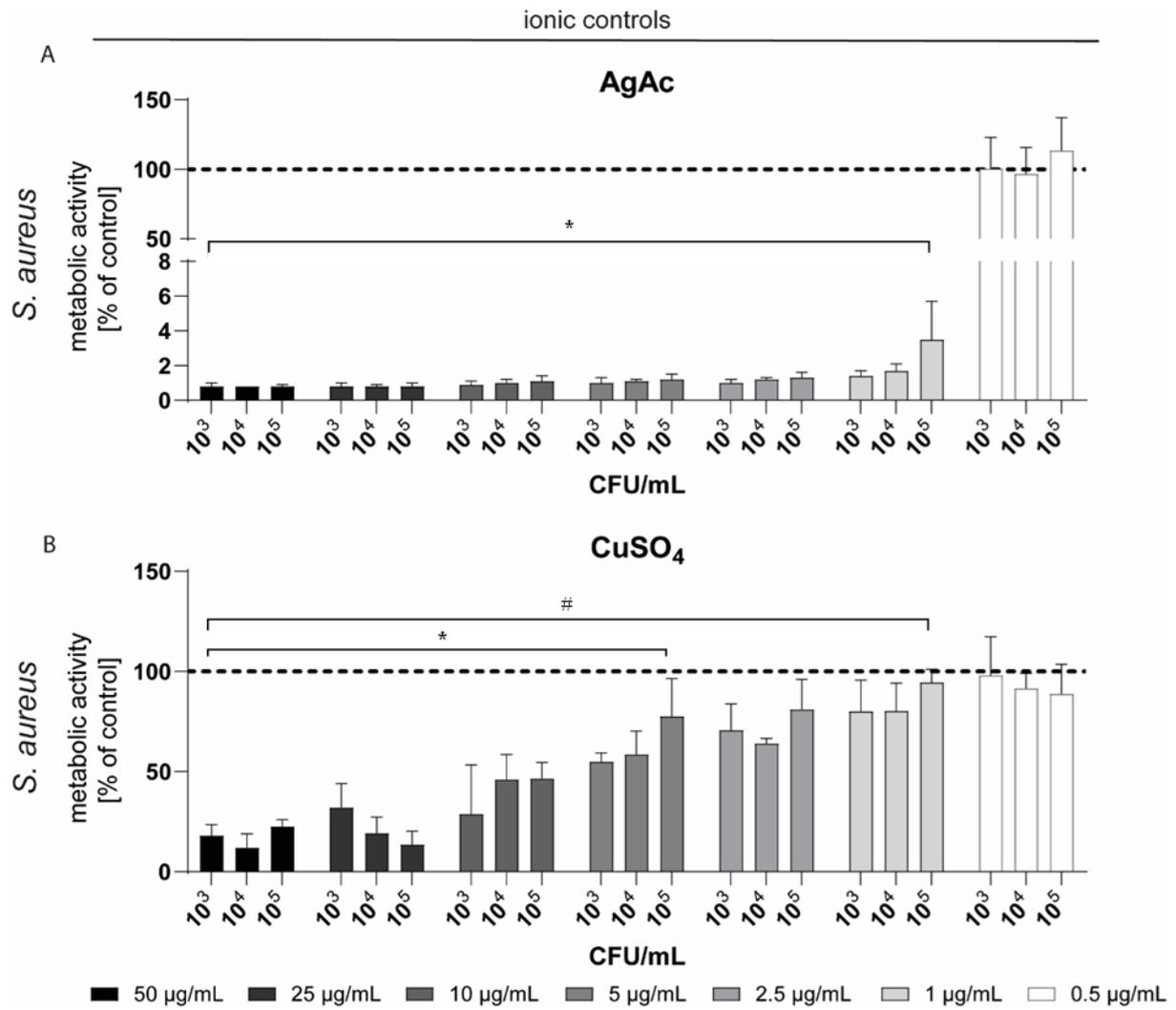

**Figure 2**: Quantitative analysis of the antibacterial activity of silver acetate (AgAc, panel A) and copper sulfate (CuSO4, panel B) solutions towards *S. aureus* (different bacterial concentrations) performed by the AlamarBlue assay. Data are expressed as mean ± SD of at least three independent experiments and given as the percentage of untreated bacteria (no exposure). Asterisks (*) indicate significant differences (* $p \leq 0.05$) compared to the untreated control; hash marks indicate significant differences (* $p \leq 0.05$) between AgAc and $CuSO_4$.

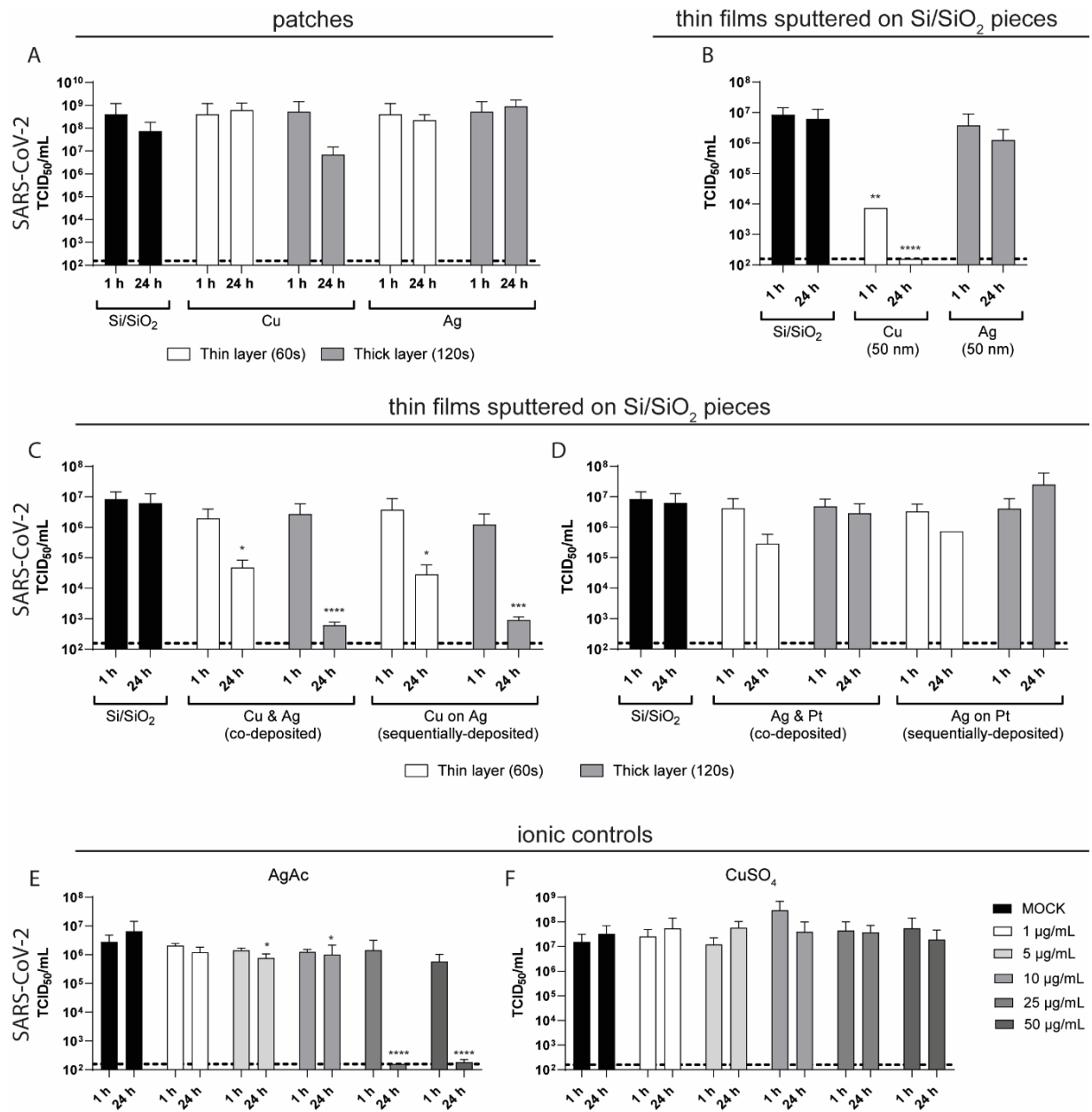

**Figure 3**: Results of antiviral activity for (A) Cu and Ag thin nanopatches sputtered on Ti and (B-D) thin films sputtered on Si/SiO$_2$ pieces which were incubated with SARS-CoV-2 for indicated time periods. (E-F) Silver acetate (AgAc) and copper sulfate (CuSO$_4$) solutions, used as ionic controls, were spiked with SARS-CoV-2 an incubated for similar time periods. Residual infectious virus was quantified by TCID$_{50}$ calculation. Dotted line indicates the lower limit of quantification. Data are expressed as mean ± SD of three independent experiments. Asterisks (∗) indicate significant differences (∗ $p < 0.05$; ∗∗ $p < 0.01$; and ∗∗∗ $p < 0.001$) compared to MOCK (untreated control) or Si/SiO$_2$.

## 4. Experimental Methods

*Sputter deposition of thin films*

Thin film samples were prepared by direct current magnetron sputtering in Ar atmosphere (0.5 Pa) at room temperature on thermally oxidized Si substrates (Si/SiO$_2$, 4.4 mm x.4.4 mm), which were placed on a rotating substrate plate. Sputter targets, 2-inch diameter, of Cu (purity 99.99%, EvoChem), Ag (99.99%, EvoChem) and Pt (99.99%, ESG Edelmetall Services) were used. Data on all films are listed in **Table 1**. The nominal thickness of the films was calculated from the pre-determined sputter rates of the used elements and the indicated power levels.

**Table 1:** Thin film surfaces prepared by magnetron sputtering.

| Element(s) | Deposition type | Power(s) [W] | Time [s] | Nominal thickness [nm] |
|---|---|---|---|---|
| Cu | | 30 | 630 | 50 |
| Ag | | 30 | 240 | 50 |
| Cu & Ag | co-deposited | Cu: 12; Ag: 5 | 60 | Cu: 2.4; Ag: 2.4 |
| Cu & Ag | co-deposited | Cu: 12; Ag: 5 | 120 | Cu: 4.8; Ag: 4.8 |
| Ag & Pt | co-deposited | Ag: 5; Pt: 5 | 60 | Ag: 2.4; Pt: 1.02 |
| Ag & Pt | co-deposited | Ag: 5; Pt: 5 | 120 | Ag: 4.8; Pt: 2.04 |
| Cu on Ag | sequential | Cu: 12; Ag: 5 | 60 | Cu: 2.4; Ag: 2.4 |
| Cu on Ag | sequential | Cu: 12; Ag: 5 | 120 | Cu: 4.8; Ag: 4.8 |
| Ag on Pt | sequential | Ag: 5; Pt: 5 | 60 | Ag: 2.4; Pt: 1.02 |
| Ag on Pt | sequential | Ag: 5; Pt: 5 | 120 | Ag: 4.8; Pt: 2.04 |
| Cu | | 12 | 60 | 2.4 |
| Cu | | 12 | 120 | 4.8 |
| Ag | | 5 | 60 | 2.4 |
| Ag | | 5 | 120 | 4.8 |

*Antibacterial tests*

Bacterial tests were performed with *Staphylococcus aureus* (*S. aureus*, DSMZ 1104) obtained from the German Collection of Microorganisms and Cell Cultures (Braunschweig, Germany). *S. aureus* cultures were grown overnight in brain-heart infusion broth (BHI broth, bioMerieux, Lyon, France) at 37 °C using a shaking water bath (JULABO GmbH, Seelbach, Germany) and bacterial concentrations were determined by turbidity measurements (Densichek turbidity photometer, bioMerieux). The adhesion and proliferation of *S. aureus* on the different nanopatch samples were analyzed using a drop-based experimental setup as reported previously [M. Köller 2015, M. Köller 2017].[10,11] Briefly, 30 µl of a bacterial solution in BHI broth containing $10^4$ cells per mL (CFU/mL) were placed in the middle of each test sample followed by incubation for 24 h in a humid chamber (water saturated atmosphere) under cell culture conditions (37 °C, 5% $CO_2$). Subsequently, the drops were aspirated, serially diluted (1:$10^4$) and plated on Columbia blood agar plates (bioMerieux) for quantitative analysis of planktonic bacteria. Qualitative analysis of the adherent bacteria was performed by SYTO-9 staining (Molecular Probes, Invitrogen, Karlsruhe, Germany) and detected by fluorescence microscopy (BX61 microscope, Olympus, Hamburg, Germany).

Silver acetate ($AgC_2H_3O_2$, AgAc) and copper sulfate ($CuSO_4$) solutions were used as ionic controls for the antimicrobial activity of Ag and Cu, respectively. Each solution was prepared in sterile ultrapure water and normalized to the total metal content (i.e., for example 100 µg/mL AgAc contains 100 µg/mL Ag). Different bacterial concentrations ($10^3$, $10^4$, $10^5$ CFU/mL) of *S. aureus* were incubated in BHI for 24 h with different concentrations of AgAc and $CuSO_4$ solutions (0.5, 1.0, 2.5, 5.0, 10, 25, 50 µg/mL) in 96-well microplates at a total sample volume of 200 µL under cell culture conditions. Subsequently, quantification of viable cells was performed by the AlamarBlue assay. Therefore, the bacterial suspensions were incubated with 20 µl of the AlamarBlue reagent (Invitrogen) until visible color change and the fluorescence intensity was analyzed at 590 nm by a microplate reader (FLUOstar Optima, BMG LABTECH GmbH, Ortenberg, Germany). The data of the treated cultures (mean ± SD) are given as percentage of the untreated controls (bacteria cultured without AgAc or $CuSO_4$).

*Antiviral tests*

To evaluate the inactivation capacity of thin films sputtered on $Si/SiO_2$ pieces (see Sputter deposition of thin films), 25 µl of SARS-CoV-2 (hCoV-19/Germany/BY-Bochum-1/2020; GISAID accession ID: EPI_ISL_1118929; $8.8 \times 10^6$ $TCID_{50}$/mL) was spotted in the middle of each test sample and incubated for 1 h and 24 h at room temperature. Virus was recovered by adding 225 µl of Dulbecco's modified Eagle's medium (DMEM, supplemented with 10 % (v/v) fetal calf serum (FCS), 1 % (v/v) non-essential amino acids, 100 IU/mL penicillin, 100 µg/mL streptomycin and 2 mM L-Glutamine). Subsequently, viral titers were determined by an endpoint dilution assay performed on Vero E6 cells (kindly provided by C. Drosten and M. Müller) seeded at $5 \times 10^4$ cells/mL in DMEM one day prior titration. The remaining $TCID_{50}$ was calculated according to Spearman and Kärber.

According to the antimicrobial tests silver acetate and copper sulfate solutions were used as ionic controls. Therefore, SARS-CoV-2 was spiked with different concentrations of AgAc and $CuSO_4$ solutions (1.0, 2.5, 5.0, 10, 25, 50 µg/mL) and was incubated for 1 h and 24 h at room temperature. Remaining viral titers were again quantified by an endpoint dilution assay followed by $TCID_{50}$ calculation.

*Statistics*

Statistical analysis of antibacterial effects was performed by one-way ANOVA with Bonferroni post hoc test. Statistical analysis of antiviral effects was performed by two-way ANOVA in a mixed-effects analysis with Dunnet's multiple comparison. p-Values ≤ 0.05 were considered as statistically significant.

**Acknowledgements**

We thank all members of the Department for Molecular & Medical Virology for helpful suggestions and discussions. E.S was supported by the VIRus ALliance NRW (VIRAL) from Ministry of Culture and Science of the State of North Rhine-Westphalia (323-8.03-151826).